\documentclass[authorversion,acmlarge]{acmart}

\AtBeginDocument{%
  \providecommand\BibTeX{{%
    \normalfont B\kern-0.5em{\scshape i\kern-0.25em b}\kern-0.8em\TeX}}}

\copyrightyear{2020}
\acmYear{2020}
\setcopyright{acmcopyright}\acmConference[CryBlock'20]{3rd Workshop on Cryptocurrencies and Blockchains for Distributed Systems}{September 25, 2020}{London, United Kingdom}
\acmBooktitle{3rd Workshop on Cryptocurrencies and Blockchains for Distributed Systems (CryBlock'20), September 25, 2020, London, United Kingdom}
\acmPrice{15.00}
\acmDOI{10.1145/3410699.3413789}
\acmISBN{978-1-4503-8079-9/20/09}

\usepackage[inline]{enumitem}
\usepackage{csquotes}
\usepackage{url}
\usepackage{hyperref}
\usepackage{multirow}

\begin{document}

\title{Are Distributed Ledger Technologies Ready for Intelligent Transportation Systems?}

\author{Mirko Zichichi}
\authornote{This work has received funding from the European Union’s Horizon 2020 research and innovation programme under the Marie Skłodowska-Curie International Training Network European Joint Doctorate grant agreement No 814177 \href{https://www.last-jd-rioe.eu/}{Law, Science and Technology Joint Doctorate - Rights of Internet of Everything}.}
\affiliation{%
  \institution{Ontology Engineering Group, Universidad Politécnica de Madrid, Spain}
}
\email{mirko.zichichi@upm.es}

\author{Stefano Ferretti}
\affiliation{%
  \institution{Department of Pure and Applied Sciences, University of Urbino "Carlo Bo", Italy} 
}
\email{stefano.ferretti@uniurb.it}

\author{Gabriele D'Angelo}
\affiliation{%
  \institution{Department of Computer Science and Engineering, University of Bologna, Italy}
}
\email{g.dangelo@unibo.it}

\renewcommand{\shortauthors}{Zichichi et al.}

\begin{abstract}
The aim of this paper is to understand whether Distributed Ledger Technologies (DLTs) are ready to support complex services, such as those related to Intelligent Transportation Systems (ITS). In smart transportation services, a huge amount of sensed data is generated by a multitude of vehicles. While DLTs provide very interesting features, such as immutability, traceability and verifiability of data, some doubts on the scalability and responsiveness of these technologies appear to be well-founded. We propose an architecture for ITS that resorts to DLT features. Moreover, we provide experimental results of a real test-bed over IOTA, a promising DLT for IoT. Results clearly show that, while the viability of the proposal cannot be rejected, further work is needed on the responsiveness of DLT infrastructures.
\end{abstract}




\maketitle

\section{Introduction}

It is widely recognized that next generation Internet services will massively resort to crowd-sourced and crowd-sensed data, coming from multiple sensors installed on multiple devices. Data aggregation provides the backbone for analyses able to capture some data findings that would not be possible from single sensors. 
This is true in Intelligent Transportation Systems (ITS) as well, where services are built through data sensed by vehicles \cite{vanderHeijden:2017}.
Transportation efficiency, travel safety, vehicle security, environment monitoring, are just few examples of types of services that might be offered \cite{mousannif2011cooperation}.
Vehicles are becoming more and more autonomous and wireless communications provide high speed connectivity, thus enabling novel smart applications.

While the amount of possible services is countless, a number of issues must be considered, that are basically related to the gathering, storing and level of trust of the data. In fact, in order to share, aggregate and trade data coming from vehicles, some features must be provided by the digital services in use, such as access control, authenticity, verifiability and proof-of-location \cite{zichichi2020distributed}. 
This is where a new kind of technology can come to aid. Distributed Ledger Technologies (DLTs) are thought to provide a trusted and decentralized ledger of data. DLTs are a novel keyword, that extends the famous ``blockchain'' buzzword, to include those technological solutions that do not organize the data ledger as a linked list of blocks. Blockchains gathered momentum when Bitcoin and other crypto-currencies skyrocketed. Then, the interest was mainly devoted to the possibility of building decentralized applications based on smart contracts \cite{D'Angelo:2018,zichichi2019like}.
Currently, DLTs are widely utilized in scenarios where: i) there are multiple parties that concur in handling some shared data, ii) there is no complete trust among these parties, and often iii) parties compete to the access/ownership of such data.
This is a typical scenario of smart transportation services that exploit data sensed from multiple sources (vehicles). Hence, the question now is if DLTs can be efficiently employed in such scenarios.

As a matter of fact, there are DLTs which have been designed with the intent to support the Internet of Things (IoT) \cite{DiPietro:2018,gda-hpcs-16,gda-simpat-iot,sf-gda}. The main features of these novel technologies are concerned with the attempt to solve some main limitations that are commonly attributed to other blockchains, such as the lack of scalability, sustainability, transaction verification rate (i.e.~how fast is the system to add novel data to the ledger).
Examples of these novel DLTs for IoT are IOTA \cite{popov2016tangle} and Radix \cite{radixkb}.
However, while their design is very interesting, at the time of writing we are aware of just few, and usually simplified, experimental studies on these technologies \cite{Bartolomeu2018IOTAFA,BROGAN2018257,Elsts:2018:DLT:3282278.3282280,zichichi2020distributed,pinjala2019}; none that demonstrate the viability of these proposed technologies in IoT and ITS scenarios.

The aim of this work is, first, to propose a novel system architecture that exploits DLTs for the support of ITS. Second, we present an experimental evaluation on DLTs, based on the use of real data traces to emulate the data generation of a smart city traffic application. We analyze the performance of the IOTA DLT, through tests that measure its degree of scalability and responsiveness in real-time scenarios.
Through our tests, we demonstrate how the Masked Authenticated Messaging (MAM) extension module of the IOTA protocol can be used to reliably and securely store and share sensed data in smart mobility applications.

Our interest is in IOTA since it that can provide permissionless access to each entity willing to participate to the ITS. It is not the scope of this work to enter into the analysis of permissioned DLTs \cite{sousa2018byzantine} because these require the preliminary creation of a consortium of trusted agents in order to govern the nodes network.

The remainder of this paper is organized as follows. Section~\ref{sec:back} provides some background on the IOTA DLT. Section~\ref{sec:model} describes the application scenario that has been built to perform the study. Section \ref{sec:perf} presents all the details of the experimental evaluation, how we conducted the experiments and which metrics have been considered. In Section~\ref{sec:res}, we describe results of the extensive experimental evaluation we conducted over IOTA. Section \ref{sec:disc} provides a discussion on the obtained results and on possible techniques to improve the DLTs performance. Finally, Section~\ref{sec:conc} provides some concluding remarks.

\section{Background}\label{sec:back}

A DLT is a software infrastructure maintained by a peer-to-peer network, where the network participants must reach a consensus on the states of transactions submitted to the distributed ledger, to make the transactions valid.
Every participant to a DLT contains a local replica of the ledger, which provides data transparency to network participants and ensures high availability of the system. 
The information recorded to a DLT is append-only, using cryptographic techniques that guarantee that, once a transaction has been added to the ledger, it cannot be modified. 

In this work, we mainly focus on IOTA, a specific DLT that is well suited for the IoT and ITS. 
This project aims to solve problems about scalability, control centralization, as well as post-quantum security issues, which are present in other blockchain technologies \cite{Bartolomeu2018IOTAFA}.
IOTA is a lightweight, permissionless DLT that enables participants to transfer immutable data and value among each other. 
From a distributed system point of view, IOTA nodes are organized as a peer-to-peer overlay, where nodes exchange messages containing updates on the decentralized ledger. 
Nodes that run the entire DLT protocol are commonly referred as full nodes.

Being the IOTA architecture still in its infancy, currently a ``coordinator node'' maintained by the IOTA foundation is present in the system. Its task is to perform a periodic checkpointing of the ledger, with the aim to sustain possible large-scale security attacks. It releases milestone transactions that confirm that all the previous transactions are valid. The purpose of the IOTA foundation is that, after the transient phase, the coordinator will be shut off, hence making IOTA a pure peer-to-peer system~\cite{coordicide}. Indeed, as shown after the ``Trinity wallet attack'' at the start of 2020 \cite{iotattack}, the IOTA foundation has the complete ability to shut down the network, simply by halting the coordinator. Clearly this is a big issue that still needs to be addressed in the IOTA development.

The IOTA decentralized ledger is not structured as a blockchain, but as a Direct Acyclical Graph (DAG) called the Tangle~\cite{popov2016tangle}. In the Tangle, graph vertices represent transactions and edges represent approvals. When a new transaction is issued, it must approve two previous transactions and the result is represented by means of directed edges. The process of attaching a novel transaction to the Tangle consists in:
\begin{itemize}
    \item \textbf{Tips selection} - selecting from the Tangle two random tip transactions that do not have a successor yet
    \item \textbf{Proof of Work (PoW)} - a computation to obtain a piece of data which satisfies certain requirements and which is difficult (costly and time-consuming) to produce but easy for others to verify \cite{popov2016tangle}. The purpose of PoW is to deter denial of service attacks and other service abuses. 
\end{itemize}

\begin{figure}[t]
  \centering
  \includegraphics[width=\linewidth]{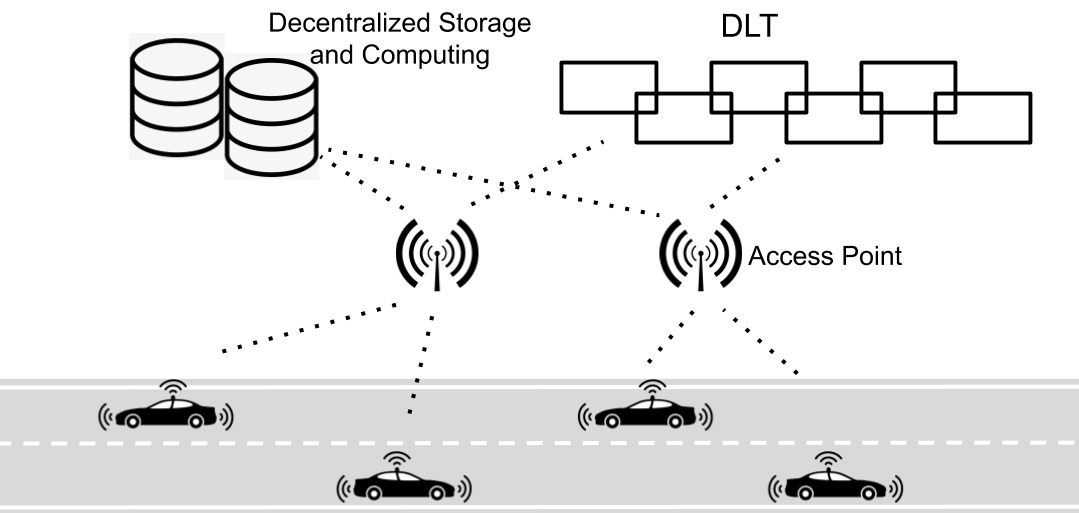}
  \caption{System model.}
  \label{fig:scenario}
\end{figure}

The validation approach is thought to address two major pain points that are associated to traditional blockchain-based DLTs, i.e.~latency and fees. IOTA has been designed to offer fast validation, and no fees are required to add a transaction to the Tangle \cite{BROGAN2018257}.
This makes IOTA an interesting choice to support smart services built through crowd-sourced data.

An important feature offered by IOTA is the Masked Authenticated Messaging (MAM). MAM is a second layer data communication protocol which adds functionality to emit and access an encrypted data stream over the Tangle. Data streams assume the form of channels, formed by a linked list of transactions in chronological order.  
Once a channel is created, only the owner can publish encrypted messages. Users that possess the MAM channel encryption key 
are enabled to decode the message. Messages are pushed on the channel in chronological order, and each message has a link to the next message to be created. Thus, once a user gains access to the MAM channel, he is enabled to see data from that moment on, whilst he cannot look back through the history of the channel before his entrance \cite{BROGAN2018257}.
In other words, MAM enables users to subscribe and follow a stream of data, generated by some devices. The data access to new data may be revoked simply by using a new encryption key.

\section{On the Use of DLTs for ITS}\label{sec:model}
We consider a set of vehicles, equipped with sensors that can generate data of some interest (see Figure \ref{fig:scenario}). 
Such sensed data can be transmitted through a network to a DLT. Thus, each vehicle interacts with a DLT node, transmitting sensed data on a periodical basis.
Data can be generally stored and manipulated by a (distributed) storage and computing platform. But the nature of this specific platform is out of the scope of this work
For instance, it might be organized as a classic cloud system, rather than a distributed file system, e.g.~IPFS \cite{Benet2014IPFSC}. 

In order to provide a level of traceability, verifiability and immutability of the generated data, the data itself, or a related digest (when the data is a large file or is sensitive information), is added to a DLT \cite{zichichi2020framework}. We assume the 
vehicle's on board computing unit
is able to issue messages to a DLT node, thanks to authentication. These messages are then converted to transactions added to the ledger. In general, all DLTs provide such kind of functionalities. For instance, in IOTA, Radix and Ethereum (e.g.~through the INFURA APIs\cite{infura}), there are APIs that allow entities, external to the DLT, to send a novel transaction.
The main point here is that these transactions must be registered in the DLT in a fast way. Second, a good level of scalability must be guaranteed. Third, since a high amount of data is produced, the DLT should offer low fees (or no costs at all). Finally, we need to treat all these transactions as a data-stream, easy to retrieve. 
By its design, IOTA is recognized as a responsive, scalable, feeless DLT, with MAM channels as the tool to treat data as streams.
For this reason, in the evaluation we will focus on IOTA.

\section{Experimental Evaluation}\label{sec:perf}

In this work, we are interested in evaluating the goodness of the adoption of IOTA as the immutable registry for ITS. Thus, we focused on the transmission of sensed data to IOTA, measuring latencies needed to issue, insert and validate transactions, and also the level of reliability of the full nodes.

\subsection{The Trace-driven Vehicles Simulation}

We conducted a trace-driven experimental evaluation.
Traces were generated using the RioBuses dataset, a real dataset of mobility traces of buses in Rio de Janeiro (Brasil) \cite{coppe-ufrj-RioBuses-20180319}. 
Based on these traces, we simulated a number of buses that, during their path, generate sensed data. 
(The type and purpose of such data is out of the scope of this evaluation, since we are mainly interested in the behaviour of the DLT; it suffices to assume that they represent typical, small sized sensed data, such as a temperatures, air pollution values, etc.)
We assume that the time spent to fetch such data is negligible, with the respect to the time to publish them to the DLT.

\begin{figure}[t]
  \centering
  \includegraphics[width=\linewidth]{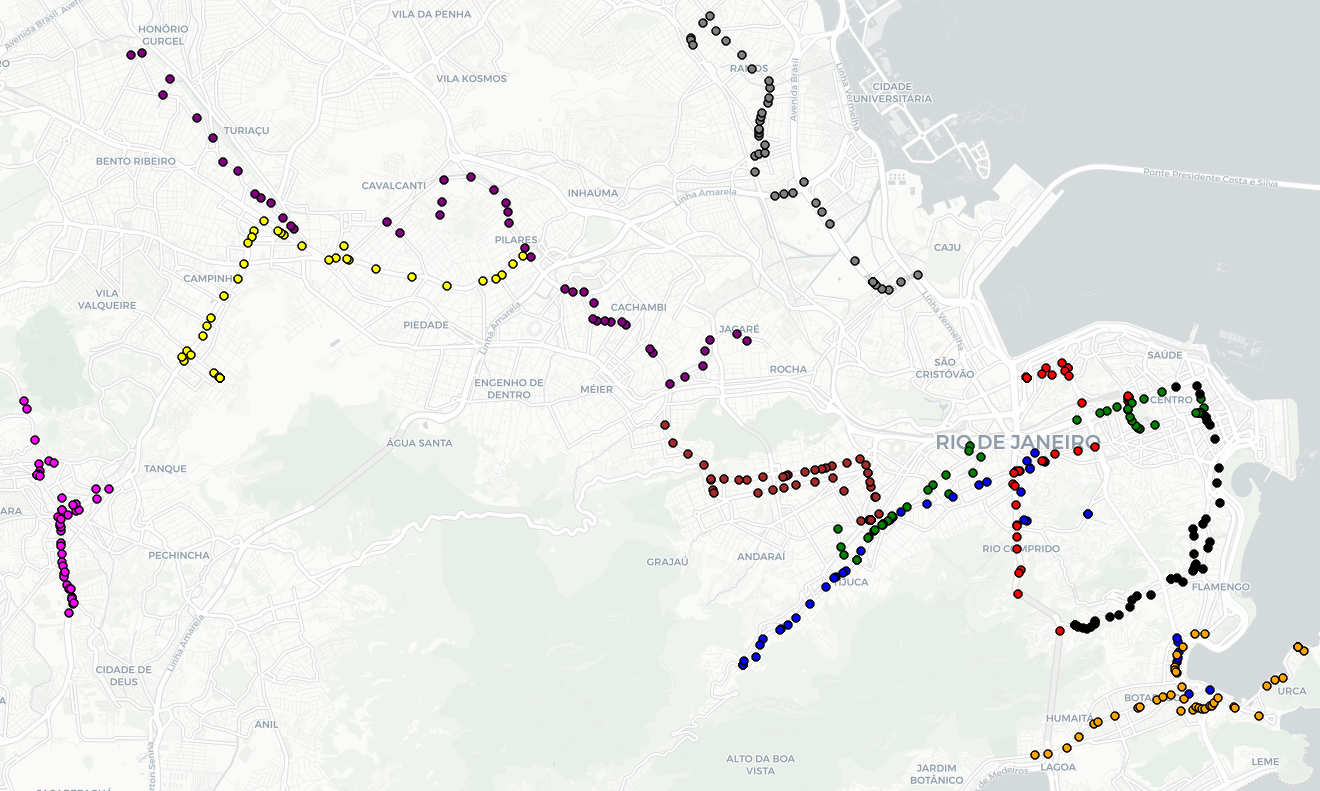}
  \caption{The 1 hour long path of 10 buses in Rio de Janeiro (Brazil).}
  \label{fig:buses}
\end{figure}

These messages were utilized to generate real transactions transmitted to the DLT. Each message was sent to a given DLT node. How this node was selected is discussed in the next subsection.
Figure \ref{fig:buses} shows the paths of 10 buses, as an example, that were considered during our tests. We varied the number of buses in the range: 60, 120, 240. 
For each bus, we utilized one hour of trace data. Based on the paths, each bus was set to generate approximately 45 message/hour. Thus, we made one hour long tests, where each bus generated, on average, a message to be issued to the DLT every 80 sec, which is a reasonable time interval to sense data in an urban scenario. For each test configuration, we replicated the experiment 12 times.

For each transaction, we recorded the outcome of the request, i.e.~successful or unsuccessful, due to some DLT nodes internal error, as well as 
the latency between the transmission of the transaction and the confirmation of its insertion in the ledger. 

\begin{figure*}
    \centering
	\includegraphics[width=\textwidth]{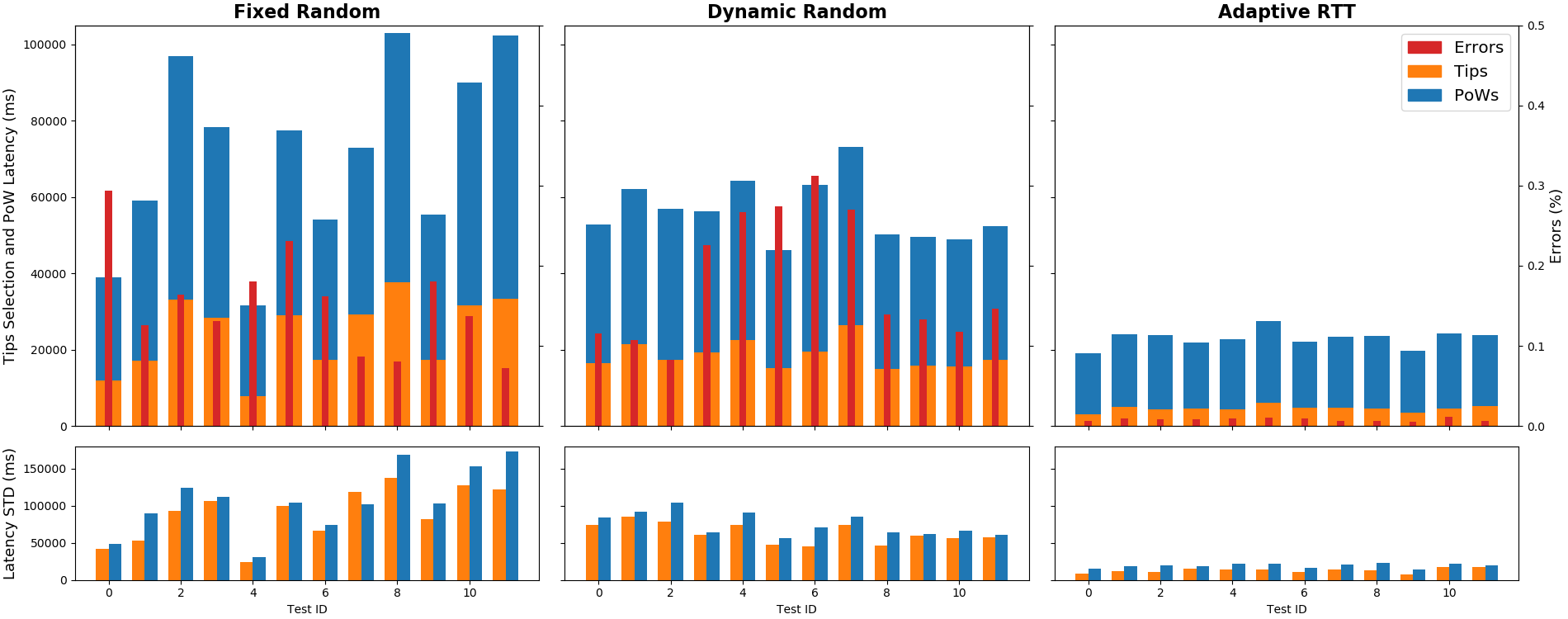}
	\caption{60 bus tests: average latencies, standard deviation and errors for the three different schemes}
	\label{fig:latencies}
\end{figure*}

\subsection{IOTA Setup}
Each bus was emulated by a single process (issuing messages based on the data trace). Thus, the first task was to find, for each bus, a full node of the IOTA DLT to interact with.
In our tests, we were enabled to rely only on services that maintain a public list of active nodes \cite{iotanodes}. This is because IOTA network full nodes do not usually allow to list their neighbors in the P2P overlay, through API. Hence, this hinders the possibility to perform a in-depth graph search on the overlay, in order to retrieve an up-to-date list of active nodes to interact with.
With this in view, the scheme we designed to select the IOTA nodes to contact for tips selection and PoW execution, is as follows. Given the list of public nodes, a filter is applied to keep only nodes that are fully synchronized, i.e.~the node has solidified all the milestones up to the latest one released by the coordinator, and that (eventually) allows remote PoW.
During testing these nodes were $\sim60$. 
Then, we designed three heuristics for the selection of a full node to pair to each bus from the public pool:
\begin{enumerate}
    \item \textbf{Fixed Random}: Each bus is assigned to a random IOTA full node from the pool, during the setup phase; then, every transaction generated by that bus is handled by this node, for the whole duration of the test.
    \item \textbf{Dynamic Random}: A random node from the pool is selected every time a message has to be published by a bus.
    \item \textbf{Adaptive RTT}: For each bus, its associated node actively changes every time a message has to be published, while the previous one is still pending. Based on results of past interactions, the known IOTA nodes are ranked through the experienced Round Trip Time (RTT) \cite{jacobson1988congestion}. Then, a new node is chosen by selecting the best known node or, if every known node is in the process of publishing a message, a new node is picked randomly from the pool. 
\end{enumerate}

We used a MAM channel associated to each single bus. 
Every message to be published in the MAM channel requires three transactions to be issued, i.e.~one containing the data and two for the signature.
The advantage of this approach is that through each MAM channel it is possible to easily retrieve the bus's data stream and that only the channel owner can publish on it. 
The entire dataset and the scripts used in this performance evaluation are stored in a GitHub repository \cite{githubrepo}.
For each transaction, we measured the time required to perform the tip selection as well as the PoW.

\section{Results}\label{sec:res}
Figure \ref{fig:latencies} shows the results obtained for different test repetitions, when the number of emulated buses was set to $60$. In particular, we show the results for each scheme we employed for the selection of the nodes. 
In the upper part, the histograms report the average latencies measured during a single test. The orange (lighter) part of the histogram shows the average latency to perform the tip selection, while the blue (darker) part shows the average latency associated to the PoW. The red (central and smaller) bars refer to the percentage of errors (the related y-axis is shown on the right of the figure), i.e.~amount of transactions that failed to be added to the Tangle, due to full nodes' errors. On the lower part of the figure, we show the average standard deviations related the specific tests, both for the tip selection and PoW. 
From the figure, it is possible to appreciate how in general a random selection of the full node to issue a transaction does not lead to good results. The amount of errors is quite high, as well as the measured latencies. Thus, these tests seem to conclude that, at the time of writing, the IOTA DLT is not fully structured to support smart services for transportation systems. 
On the other hand, the good news is that if we carefully select the full node to issue a transaction, the performances definitely improve. In fact, our third scheme ``Adaptive RTT'' has a low amount of errors, on average around 0.8\%. Measured latencies are lower than other approaches because well performing nodes are chosen more often. Still, the average latency amounts to 23 seconds, which is far from a real-time update of the DLT. The level of acceptability of latency values truly depends on the application scenario. 

\begin{figure}
    \centering
	\includegraphics[width=.45\textwidth]{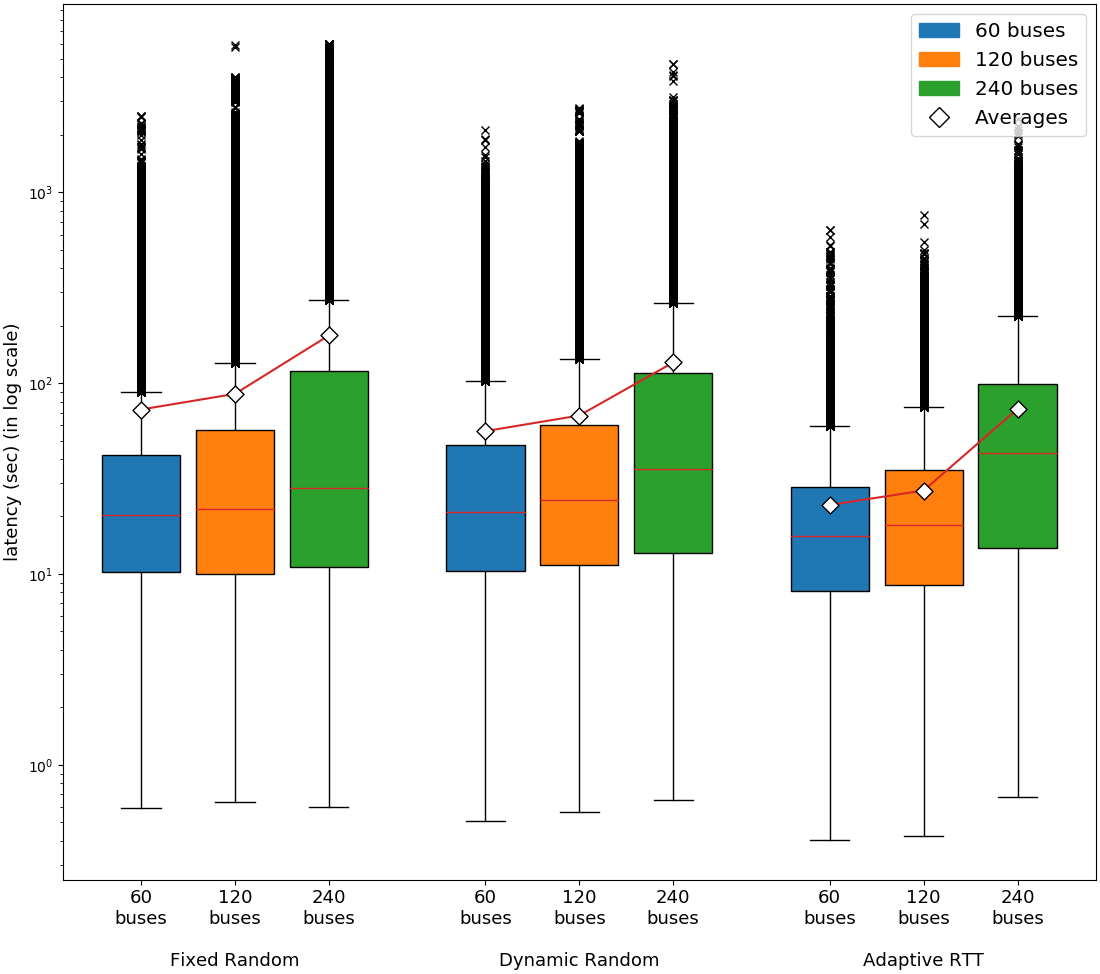}
	\caption{Boxplots for tests with 60, 120, 240 buses (y-axis in log-scale)}
	\label{fig:boxplot}
\end{figure}

These first results suggest that some scalability tests might give further insights on the viability of the use of IOTA as the DLT to support ITS. For this reason, we made some tests with an increasing number of buses.
Figure \ref{fig:boxplot} shows average results obtained using our three considered schemes, when varying the number of buses. Results are reported as box plots. Thus, each box plot corresponds to the average results for a scheme in a given scenario. This allows to assess the scalability of each scheme, by looking at the results for an increasing amount of buses. At the same time, it is possible to compare the three schemes by looking at their performance for each scenario.

In the box plot, the diamond represents the mean value of the overall latency in transaction transmission (i.e. tips selection + PoW). 
The rectangle identifies the Inter-Quartile Range (IQR), i.e.~values from the 25th to the 75th percentile, representing the middle 50\% of values. 
Hence, the lower part of the box (let denote it Q1) is the first quartile (25th percentile), the highest (denote it Q3) is the third quartile (75th percentile). 
The red line inside the box is the median value. The lower and upper values identified by the vertical line are the whiskers. 
In box plot, the whiskers are defined as 1.5 times the IQR.
Thus, the lower whisker is Q1 - 1.5*IQR, while the upper whisker is Q3 + 1.5*IQR; they represent a common way to describe the dispersion of the data. Finally, the ``$\times$'' symbols outside the whiskers are the outliers.
To better show the obtained results, the y-axis is reported in a log scale.

Results show that in all cases, average latencies increase significantly with the number of buses. It is worth noticing that, being the y-axis in log scale, the difference on the performance is relevant.
It is also confirmed that ``Adaptive RTT'' provides better results since average latencies are definitely lower than other schemes. In particular, the first two schemes have outliers well over $10^3$ sec.
This suggests that the number of full nodes devoted to the transaction management should increase proportionally to the number of buses. 

In case 240 buses are active and send requests, we know that about $\sim 3$ msg/sec must be stored in IOTA. Assuming that the workload is evenly distributed among 60 nodes, then, each node should provide for 4 buses, receiving, on average, a new transaction request every $ \sim 20$ seconds.
Bearing in mind that at best it takes 23 seconds (on average, using "Adaptive RTT") for a full node to process a transaction, then we see that an initial overhead of a few seconds leads to a huge increase at the end of the test.
It is worth noticing, however, that in "Adaptive RTT" test cases the same $\sim 15$ nodes were always requested (as they performed better than the others), with an average workload of 16 buses per node, which means a new request every 5 seconds. For this reason the latency increased to 73.26 seconds on average.
This means that further improvements are needed to solve scalability issues.

\begin{table*}[htb]
\centering\caption{Results on IOTA, with 60, 120, 240 buses.\label{tab:iota}}
    \begin{tabular}{|c|c|c|c|c|}
        \hline
        \hline
        \# \textbf{buses} & \textbf{Heuristic} & \textbf{Avg Latency} & \textbf{Conf. Int. (95\%)} & \textbf{Errors}\\ 
        \hline
        \multirow{2}{*}{60} & Fixed Random & 72.68 sec & [70.43, 74.94] sec & 15.37\% \\
        \cline{2-5}
         & Dynamic Random & 56.0 sec & [54.51, 57.5] sec & 18.26\% \\
        \cline{2-5}
         & Adaptive RTT & 22.99 sec & [22.69, 23.29] sec & 0.81\% \\
        \hline
        \hline
        \multirow{2}{*}{120} & Fixed Random & 87.75 sec & [85.38, 90.12] sec & 29.49\% \\
        \cline{2-5}
         & Dynamic Random & 67.6 sec & [66.29, 68.9] sec & 18.99\% \\
        \cline{2-5}
         & Adaptive RTT & 27.35 sec & [27.11, 27.58] sec & 1.1\% \\
        \hline
        \hline
        \multirow{2}{*}{240} & Fixed Random & 177.62 sec & [174.25, 181.0] sec & 42.81\% \\
        \cline{2-5} 
        & Dynamic Random & 128.2 sec & [126.28, 130.12] sec & 44.85\% \\
        \cline{2-5} 
        & Adaptive RTT & 73.26 sec & [72.68, 73.85] sec & 7.55\% \\
        \hline
        \hline
    \end{tabular}
\end{table*}

To better emphasize the outcomes, Table \ref{tab:iota} 
reports some summarized statistics (shown in the box plots) and the error rates. Actually, another great difference in the performance of the approaches is in the amount of errors. While the average error for ``Adaptive RTT'' is $\sim1$\%, for the other two schemes we have errors well above $15$\%. These error rates are clearly unacceptable, meaning that these approaches are unusable.

\begin{figure}
    \centering
	\includegraphics[width=.45\textwidth]{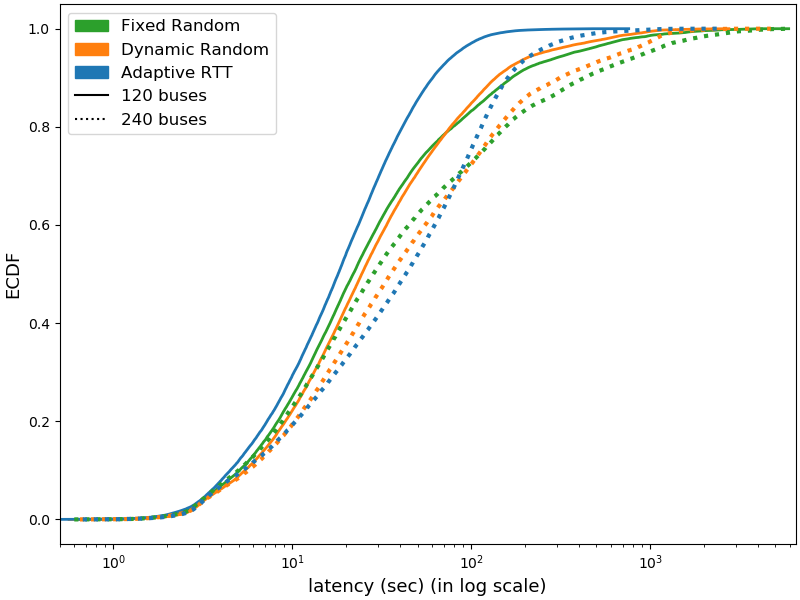}
	\caption{Empirical Cumulative Distribution Function for tests with 120 and 240 buses (x-axis in log-scale)}
	\label{fig:ecdf}
\end{figure}

Finally, Figure \ref{fig:ecdf} shows the empirical cumulative distribution function obtained for the compared schemes in the 120 and 240 bus scenarios. In this case, for the sake of a better visualization, the x-axis is shown in a log-scale. These charts further confirm the better performance obtained by the ``Adaptive RTT'' scheme.

\section{Discussion}\label{sec:disc}

Obtained results require a discussion. In fact, on one side, it is shown that through a proper selection of full nodes it is possible to achieve reliable ledger updates (low errors), thus making viable the use of IOTA to support ITS. On the other side, however, the measured latencies are relevant.
In our tests, we employed public available IOTA full nodes to add transactions. Thus, we refer to such nodes to perform the tip selection and the PoW. The rationale behind this choice was based on the assumption that buses' computing units do not have computation capabilities to behave as full nodes \cite{Elsts:2018:DLT:3282278.3282280}.

During our experimental evaluation, all the full nodes had usually a low computational load. Nonetheless, results confirm that the selection of the node is quite relevant. As a further confirmation of this claim, in our preliminary tests we tried to exploit a heuristics, alternative to those presented in the previous section. The idea was to select the best N full nodes, in terms of available resources, and use them to issue the transactions. 
We found out that knowing \textit{a priori} the best full nodes is not feasible and that it was simpler (with similar performances) employing the ``Fixed Random'' approach. 
For this reason an algorithm like the ``Adaptive RTT'' is needed to find best nodes at run time.

An alternative approach might be to employ an edge computing system model, where the execution of the PoW is executed locally in a gateway. In this case the tip selection must be always accomplished at a full node, that maintains a complete copy of the Tangle. The rationale would be to relieve the IOTA node from the computational burden of the PoW. However, this would force to equip the gateway with sufficient computational capabilities to perform the PoW for all the transactions generated by the buses it handles.

Finally, it would be possible to ask the gateway to act as a full node for the DLT. This would actually resemble the testbed we considered in this work (due also to the fact that the exploited IOTA public nodes had a low workload overhead, concurrent to our tests). In this case, the difference is that it would be possible to have a direct control of the full node. Its hardware characteristics might be properly set to tolerate a certain predicted workload, and this node might be reserved to handle transactions from the specific ITS application, only.
These two scenarios represents an interesting further work.

\section{Conclusions}\label{sec:conc}

In this paper, we proposed an architectural solution resorting to DLTs to support ITS. The benefits on the use of the distributed ledgers is that they would allow to safely and securely store sensed data, offering authenticity, verifiability and immutability features. Moreover, the use of DLTs can be employed to provide proof-of-location \cite{zichichi2020framework}. 

We analyzed the main characteristics of current permissionless DLTs and focused on the DLT that, among the others, promises to be the best solution for intelligent transportation scenarios, i.e.~IOTA. We thus made an extensive experimental evaluation, whose results have been summarized and analyzed. The conclusion is that, probably, work must be done, in order to provide effective distributed ledgers for ITS. In fact, it is important to be able to select proper nodes to interact with in order to have acceptable error rates. Moreover, measured latencies resulted higher than 20 sec, which is quite high if we think at real-time applications, reasonable for less time demanding services. In any case, this might be a transient problem, that could be solved by improving the IOTA peer-to-peer infrastructure.

Furthermore, in our tests all the work (i.e.~tip selection and PoW) was performed by the full nodes. The rationale was to relieve sensors and devices from this task \cite{Elsts:2018:DLT:3282278.3282280}.
An alternative solution might be to delegate the PoW to some other entity, such as a gateway in between the vehicle sensors and the full node.
Moving the PoW from the full nodes elsewhere might strongly improve the performances of the DLT nodes. The study of this possible improvement is ongoing.

\bibliographystyle{ACM-Reference-Format}
\bibliography{paper}

\end{document}